\documentstyle[prc,preprint,aps,amstex,amssymb]{revtex}
\begin{document}
\preprint{}
\tighten
 
\title{The $\boldsymbol{\pi^+}$--emission puzzle
       in $\boldsymbol{^4_\Lambda}$He decay}

\author{B.\ F.\ Gibson}
\address{Theoretical Division, Los Alamos National Laboratory, \\
         Los Alamos, NM 87545, USA}
\author{R.\ G.\ E.\ Timmermans}
\address{Kernfysisch Versneller Instituut, University of Groningen, \\
         Zernikelaan 25, 9747 AA Groningen, The Netherlands}

\date{\today}

\maketitle

\begin{abstract}
We re-examine the puzzling $\pi^+$ emission from the weak decay
of $^4_\Lambda$He and propose an explanation in terms of
a three--body decay of the virtual $\Sigma^+$. Such a resolution
of the $\pi^+$ decay puzzle is consistent with the calculated
$\Sigma^+$ probability in light $\Lambda$ hypernuclei as well
as the experimentally observed $\pi^+$ energy spectrum and
$s$--wave angular distribution.
\end{abstract}

\section{INTRODUCTION}
The observed~\cite{May66,Key76} $\pi^+$ emission from the weak decay
of $^4_\Lambda$He has been an intriguing puzzle for more than 30 years.
Experimentally, the $\pi^+$ to $\pi^-$ ratio for $^4_\Lambda$He decay
is about 5\%, whereas no unambiguous $\pi^+$ events have been observed
for other hypernuclei. Since the mesonic decay modes of the free
$\Lambda$ ($\to p+\pi^-$, $n+\pi^0$) produce only $\pi^-$s and
$\pi^0$s, more complicated mechanisms must be responsible for
the $\pi^+$s in the decay of the bound $\Lambda$ in $^4_\Lambda$He.
In this paper, we propose a solution to the puzzle involving the
virtual $\Sigma^+$ component of the $^4_\Lambda$He wave function.

The strangeness $-1$ $\Lambda\leftrightarrow\Sigma$ conversion, which
leads to strong coupling of the $\Lambda N$ and $\Sigma N$ channels,
appears to play an important role in the physics of the light
$\Lambda$ hypernuclei.  The $\Lambda N-\Sigma N$ coupling is
much more significant in hypernuclei than is the (octet--decuplet)
coupling of the nucleon to the $\Delta$(1232) in ordinary nuclear
physics. Not only is the $\Sigma$ stable with respect to the strong
interaction, whereas the $\Delta$ is not, and the mass difference
between $\Lambda$ and $\Sigma$ just some 80 MeV, the octet--decuplet
coupling in the nonstrange sector ($N\!N-\Delta N$) can, because of
duality, be at least partially subsumed in the one--boson--exchange
(OBE) model of the $N\!N$ interaction, without requiring explicit
$\Delta$s.

The few--body $\Lambda$ hypernuclei data clearly suggest that 
$\Lambda N-\Sigma N$ coupling is important. The hypertriton, 
$^3_\Lambda$H, would likely not be bound without it~\cite{Afn90,Glo95}.
Charge symmetry breaking (CSB) in the $A=4$ isodoublet is obvious:
$^4_\Lambda$He is more bound than $^4_\Lambda$H by 0.35 MeV,
almost three times the 0.12 MeV estimated for the $^3$H--$^3$He
isodoublet once the Coulomb energy is taken into account. In other
words, the $\Lambda\,p$ interaction is stronger than the $\Lambda\,n$
interaction. This CSB in the hyperon--nucleon ($Y\!N$) interaction
arises in large part due to the $\Sigma^+-\Sigma^-$
mass difference of some 8 MeV (about 10\% of the 80 MeV
$\Lambda-\Sigma$ mass difference); $\Lambda\,p$ couples to
$\Sigma^+\,n$, while $\Lambda\,n$ couples to $\Sigma^-\,p$.
Furthermore, significant $\Lambda-\Sigma$ conversion effects
in $\Lambda$ hypernuclei are also suggested by comparing
$\Lambda$ separation energies from $^3_\Lambda$H,
$^4_\Lambda$He, and $^5_\Lambda$He with neutron separation
energies from $^2$H, $^3$H, and $^4$He. The ratio of neutron
separation energies for neighboring $s$--shell nuclei is close
to a constant:
\begin{eqnarray*}
  B_n(^3{\rm H})/B_n(^2{\rm H})   & \simeq & 6/2  = 3      \ , \\
  B_n(^4{\rm He})/B_n(^3{\rm H}) & \simeq & 20/6 \simeq 3 \ .
\end{eqnarray*}
Therefore, one might anticipate for the $\Lambda$ separation energy
that 
\begin{equation}
  B_\Lambda(^5_\Lambda{\rm He}) \simeq 3 \,
  B_\Lambda(^4_\Lambda{\rm H}) \simeq 6 \; {\rm MeV} \ , \nonumber
\end{equation}
for calculations in which $V_{\Lambda N}$ is fitted to the
low--energy $\Lambda N$ scattering data and which reproduce the
observed value of $B_\Lambda(^4_\Lambda{\rm H})=2.04(4)$ MeV. This
is confirmed by model calculations~\cite{Bod66,Her67,Gib72,Gal75} 
in which central potentials represent the $\Lambda N$ force.
But such is not the case experimentally; the observed value for 
$B_\Lambda(^5_\Lambda$He) is only 3.10(2) MeV~\cite{Jur73,Dav67,Dav91}.  
Similarly, for the hypertriton, one might anticipate that
\begin{equation}
  B_\Lambda(^3_\Lambda{\rm H}) \simeq \frac{1}{3}
  B_\Lambda(^4_\Lambda{\rm H}) \simeq 0.7 \; {\rm MeV} \: . \nonumber
\end{equation}
This is the model result using central potentials to represent
the $\Lambda N$ interaction~\cite{Afn90,Dab73,Gib75}. However,
the experimental value for $B_{\Lambda}(^3_{\Lambda}$H) is 0.13(5)
MeV. From these considerations one can infer that $\Lambda N-\Sigma N$
coupling is an important aspect of modeling the $Y\!N$ interaction.

Until now a cogent explanation of the $\pi^+$ decay of $^4_\Lambda$He 
has been lacking. Dalitz and von Hippel~\cite{Dal62,Dal64,Hip64}
explored the issue in depth. They considered two--body decay
processes of the type: ($i$) $\Lambda\to\pi^0+n$ decay followed
by a $\pi^0+p\to\pi^++n$ charge--exchange reaction, and ($ii$)
$\Sigma^+\to\pi^++n$ decay following a $\Lambda+p\to\Sigma^++n$
conversion. Their conclusion was that neither
process could account for more than a 1\% $\pi^+$ decay rate.
Dalitz discussed the possibilities at length in his Varenna
lectures~\cite{Dal66} but apparently found the problem intractable.
He argued that the experimental observation identifying
$\Sigma^+\to\pi^++n$ decay as a $p$--wave process ruled
out the promising explanation coming from von Hippel's
calculations~\cite{Hip64}, which had found that only $s$--wave $\Sigma^+$
decay might yield a sufficiently high rate. More recently, Ciepl\'y
and Gal~\cite{Cie97} have re-examined the charge--exchange contribution
to $\pi^+$ emission in $^4_\Lambda$He decay. They concluded that
even though their improved calculation with up--to--date input
parameters yields a 1.2\% branching ratio, some twice as large as
that obtained by Dalitz and von Hippel, the charge--exchange mechanism
by itself cannot account for the experimental value of about 5\%.

The remainder of this paper is organized as follows.
In Sec.\ II we review the experimental situation which leads to
our suggested explanation for the $\pi^+$ decay mode. In Sec.\ III 
we consider the theoretical model assumptions and estimates. Finally,
our conclusions are summarized in Sec.\ IV.

\section{EXPERIMENTAL DATA}
The ratio of $\pi^+$ decays to $\pi^-$ decays for $^4_{\Lambda}$He
is defined as
\begin{equation}
  R(\pi^+/\pi^-) =
     \frac{\Gamma(^4_\Lambda\rm{He}\to\:all\ \pi^+\: modes)}
          {\Gamma(^4_\Lambda\rm{He}\to\:all\ \pi^-\: modes)} \: .
\end{equation}
The measurement coming from the bubble chamber study by Keyes
{\it et al}.~\cite{Key76} yielded a value $R(\pi^+/\pi^-)=4.3(1.7)\%$.
Results from Mayeur {\it et al}.~\cite{May66} and from Bohm
{\it et al}.~\cite{Boh69} are quoted as lying within the range
\begin{equation}
    5.4^{+1.5}_{-1.7}\% \:\leq\:R(\pi^+/\pi^-)\:\leq\: 
    6.9^{+1.8}_{-2.1}\% \: . \nonumber
\end{equation}
Thus, we see an approximately 5(2)\% $\pi^+$ decay probability
observed in the experiments.

Sacton's review~\cite{Sac66} of the experimental situation provides a
cogent summary of these results, including those from the papers by Mayeur
{\it et al}.~\cite{May66} and by Gajewski {\it et al}.~\cite{Gaj66}.
We reproduce the data in Fig.\ 1 in which we compare the pion kinetic
energy spectra for the following four decay processes:
\begin{eqnarray*}
  ^4_\Lambda{\rm He} &\to& \pi^- +p+\,^3{\rm He} \hspace{4mm} (a) \\
  ^4_\Lambda{\rm He} &\to& \pi^+ +n+\,^3{\rm H}  \hspace{6mm} (b) \\
  ^4_\Lambda{\rm H}  &\to& \pi^- +n+\,^3{\rm He} \hspace{4mm} (c) \\
  ^4_\Lambda{\rm H}  &\to& \pi^- +p+\,^3{\rm H}  \hspace{6mm} (d)
\end{eqnarray*}
The $\pi^-$ spectrum from $^4_\Lambda$He into $p +\,^3$He, process 
($a$), is peaked at a kinetic energy of around 30 MeV, as one would
expect for $\pi^-$s coming from an underlying $\Lambda\to\pi^-+p$
free decay. The tail extends down to 15 MeV. The $\pi^-$ decay
of $^4_\Lambda$H into $p +\,^3$H, process ($d$), is similarly
peaked but several MeV higher.  The primary strength for
$^4_\Lambda$H decay lies in the $^4_\Lambda$H $\to\pi^-+\,^4$He
analog mode~\cite{Boh69}, so that the spectrum of ($d$) contains many
fewer events than that of ($a$). Krecker {\it et al}.~\cite{Kre74}
present a later summary of such $\pi^-$ decay data. Our interpretation 
of ($a$) is that one is looking at processes dominated by
$\Lambda\to\pi^-+p$ decay embedded within a very light nucleus, so 
that Fermi smearing of the peak is limited. The low--energy side
of the peak could easily come from final--state rescattering. The
possible three--body $\Lambda+N\to\pi^-+p+N$ decay process is less 
likely to leave behind a bound trinucleon.  Indeed, specific decay 
events involving two protons in the final state have been 
identified~\cite{Boh69}.  However, we would expect to see $\pi^-$ 
events with kinetic energies of less than 15 MeV, if $\pi^-$s from 
three--body decay processes were of consequence.

Perhaps surprisingly, the $\pi^-$ kinetic energy spectrum for
$^4_\Lambda$H into $n +\,^3$He, process ($c$), and into $p +\,^3$H, 
process ($d$), look similar to that for ($a$).  They exhibit a peak 
in the region corresponding to the $\Lambda\to\pi^-+p$ free decay 
peak. (In addition, there exist four events at low energy, below 15 
MeV.)  We suggest that the two--body ($\Lambda\to\pi^-+p$) decay 
appearance comes from
\begin{equation}
 ^4_\Lambda{\rm H} \to \pi^-+\,^4{\rm He}^* \hspace{4mm} (e) \nonumber
\end{equation}
where the $^4$He$^*$ $T=0$ states decay equally into $n +\,^3$He
and $p +\,^3$H. Sacton shows in his Table I that the
two--body decay $^4_\Lambda$H $\to \pi^- +\,^4$He generates
some ten times the number of events that the three--body decay
$^4_\Lambda$H $\to \pi^- + p +\,^3$H produces. Thus, we infer ($i$)
that the three--body decay modes ($c$) and ($d$) are closely related,
as the number of events for each in Fig.\ 1 indicates, and ($ii$)
that they come primarily from the decay of the $T=0$ $^4$He excited
states, following $\pi^-$ emission. These conclusions about the strong
final--state interactions involved in ($c$) and ($d$) are supported
by the study in Ref.~\cite{Ada73}, where it is argued that a naive
calculation not taking into account resonant final states fails.
However, the low--energy $\pi^-$s are most likely to come from
multi--nucleon final states, as we discuss below.

Such a picture is not in contradiction with the physics of process
($a$), where the lowest threshold for $\pi^-$ decay of $^4_\Lambda$He
is through the $T=1$ excited states of $^4$Li. The $T=1$ four--nucleon
interactions are much weaker than the $T=0$ interactions. Thus, the
nucleonic final--state interactions should be of less consequence in
process ($a$), which leads one to a spectrum for the observed $\pi^-$s
more in agreement with naive expectations for a decay dominated by the
free $\Lambda\to\pi^-+p$ process with no final--state interactions.
Moreover, the peak in ($a$) should be at lower energy than that
in ($c$) and ($d$), as it indeed appears to be in Fig.\ 1.

The $^4_\Lambda$He $\to \pi^+ + n +\,^3$He decay mode ($b$) is the
puzzle that we wish to address. We observe from Fig.\ 1 that, unlike
the $\pi^-$ decay spectra which are peaked according to two--body decay
($\Lambda\to\pi^-+p$) expectations, the $\pi^+$ spectrum from the
$^4_\Lambda$He $\to\pi^+$ decay is flat in terms of the $\pi^+$ energy
distribution. Moreover, it is stated in Ref.~\cite{May66} that for
$\pi^+$ kinetic energies below 22 MeV multi--neutron final states
are likely. Therefore, the label ``$^3$H'' in process ($b$) and
in the caption for Fig.\ 1$b$ should be interpreted as
``$^3$H or $n+\,^2$H or $n+n+\,^1$H.''
This is even more clearly stipulated in Ref.~\cite{Key76}, where the
final state is labeled ``$nnnp$.'' Furthermore, we note the paucity
of events for pion kinetic energies above that corresponding to the
threshold for four--nucleon decay. Analogously, the low--energy
$\pi^-$s seen in the $\pi^-$ spectra in Fig.\ 1 are likely due to such
multi--nucleon final states. The more complete compilation of $\pi^+$
decay data from Ref.~\cite{Boh69}, as reproduced in Fig.\ 2, confirms
the flat character of the $\pi^+$ spectrum. Moreover, Keyes
{\it et al}.~\cite{Key76} argue that the data suggest the $\pi^+$
emission process is predominantly $s$--wave. Therefore, we conclude
there is no evidence that the two--body decay of the virtual
$\Sigma^+$, assumed to be operative by von Hippel, can account for
the $\pi^+$ spectrum.

\section{THEORETICAL ASPECTS}
Dalitz and von Hippel considered a number of second--order processes
to explain the $\pi^+$ emission of $^4_\Lambda$He. The possibility
of charge exchange $\pi^0+p\to\pi^++n$ following $\Lambda\to\pi^0+n$
decay was estimated~\cite{Dal64} to provide at most a 0.6\%
decay rate probability. Furthermore, Dalitz suggested
that $\Lambda+p\to\pi^-+p+p$ and $\Lambda+n\to\pi^-+n+p$
decays, considered in Ref.~\cite{Dal62}, do not appear to contribute
significantly to the $\pi^-$ spectra for pion kinetic energies below
15 MeV, where a significant number of the $\pi^+$ events lie.
Hence, $\pi^+$ decay would necessarily need to occur via a different 
three--body mechanism, if it is not to be ruled out by the $\pi^-$ 
decay spectrum covering the region of pion kinetic energies expected 
to be dominated by three--body decay processes.

It was such a virtual process
($\Lambda+p\to\Sigma^++n\to\pi^++n+n$) that von Hippel studied.
However, von Hippel estimated~\cite{Hip64} that $\Sigma^+\to\pi^++n$
decay contributes at most about 0.2\% to the $\pi^+$/$\pi^-$ ratio,
assuming the decay $\Sigma^+\to\pi^++n$ proceeds preferentially
through a relative $p$--state of the $\pi^+ n$ system, as has been 
observed experimentally for free $\Sigma^+$ decay.  Von Hippel's 
numerical estimate was based upon a median $\pi^+$ kinetic energy 
of 10 MeV, in his closure approximation.  The mean $\pi^+$ kinetic 
energy is more like 18 MeV, which should raise von Hippel's estimate 
by a factor of about two, still too small to account for the 
experimental observations.

Nevertheless, it is the large probability for a virtual $\Sigma^+$
which is unique to the $^4_\Lambda$He hypernucleus: The wave
function of $^4_\Lambda$He can be written schematically as
\begin{equation}
   |\,^4_\Lambda{\rm He}\rangle =
 \alpha\,|\Lambda\!\otimes\!\,^3{\rm He}\rangle + \beta\,(
  -\sqrt{\frac{1}{3}}\,|\Sigma^0\!\otimes\!\,^3{\rm He}\rangle
  +\sqrt{\frac{2}{3}}\,|\Sigma^+\!\otimes\!\,^3{\rm H}\rangle ) \ .
\end{equation}
Because of its charge, $^4_\Lambda$He does permit the $\Lambda$
to make a virtual transition to a $\Sigma^+$ without altering
the structure of the ``nuclear core'' state, whereas one would
anticipate only $\Sigma^-$ (and $\Sigma^0$) transitions in
$^4_\Lambda$H and other $\Lambda$ hypernuclei.
This suggests that the $\Lambda+p\to\Sigma^++n$ transition
is the key to understanding the $\pi^+$ emission.

With this in mind, we interpret the flat $s$--wave $\pi^+$ spectrum
seen above as evidence for a {\it three--body} decay mechanism of
the type $\Sigma^++N\to\pi^++n+N$
replacing the $\Sigma^+\to\pi^++n$ ``free'' decay unavailable to
the deeply bound $\Sigma$. 
The virtual $\Sigma^+\,N$
system is ``off--shell,'' and hence there must be a
$\Sigma^++N\to\pi^++n+N$ rescattering reaction to restore
the system to ``on--shell'' and free the observed $\pi^+$s.
We expect both nucleons to carry off kinetic energy, producing a
pion spectrum more--or--less uniformly distributed over the allowed
energies from zero to the maximum corresponding to a $n+n+\,^2$H final
state. One of the two neutrons in the $\pi^+nn$ rescattering process
could be ``picked up'' by the spectator deuteron to produce a triton. 
Alternatively, the proton in the $\pi^+np$ rescattering process
could be picked up by the spectator di-neutron to produce a triton.
However, were the $^3$H final state to play a dominant role in the
$\pi^+$ decay, one would expect to see primarily $\pi^+$ events above
the $nn\,^2$H threshold in Fig.\ 1$b$. A three--body decay amplitude 
of the $\Sigma^++N\to\pi^++n+N$ type, normalized to the theoretically
estimated $\Sigma^+$ probability in $^4_\Lambda$He, can explain not
only the $\pi^+$ decay branching ratio but also the $s$--wave angular
distribution and the flat energy distribution of the $\pi^+$s.

One simple estimate of the $\Sigma^+$ probability in $^4_\Lambda$He
can be obtained as follows. Gl\"ockle and coworkers~\cite{Glo95}
have calculated the $\Sigma$ probability for the very weakly bound
hypertriton to be 0.5\%. Given the small $^3_\Lambda$H binding
energy of 0.13(5) MeV, the $\Sigma$ probability $|\beta|^2$ for
$^4_\Lambda$He with $\Lambda$ separation energy of 2.39(3) MeV
will be much larger.  For example, using a linear extrapolation,
\begin{equation}
  P(\Sigma) = |\beta|^2 \simeq
              2.39(3)/0.13(5)\times 0.5\% = 9(3)\% \ ; \nonumber
\end{equation}
Taking into account the 3/2 for spin--1 $\Lambda p$ pairs in
$^4_{\Lambda}$He and the 1/4 for such pairs in $^3_\Lambda$H
would push this significantly higher. Alternatively, this $\Sigma$
probability has been estimated in a model calculation~\cite{Gib72}
for the $^4_\Lambda$He--$^4_\Lambda$H system to be as large as 14\%.
Therefore, we assume the $\Sigma^+$ probability in $^4_\Lambda$He
to be
\begin{equation}
  P(\Sigma^+) = \frac{2}{3}\,|\beta|^2 =
                \frac{2}{3}\times 14(6)\% = 9(4)\% \ .
\end{equation}

In order to translate this $\Sigma^+$ probability into an
estimate for the branching ratio $R(\pi^+/\pi^-)$, we need
to take into account suppression effects due to the Pauli
principle for the in--medium decay rates, normalized to
the $\Lambda$ and $\Sigma^+$ lifetimes in vacuum.
According to Dalitz and Liu~\cite{Dal59}, the $\pi^-$ decay
rate of $^4_\Lambda{\rm He}\to\pi^-+p+\,^3$He, with three
protons in the final state, is strongly Pauli suppressed
with respect to the $\Lambda\to\pi^-+p$ decay rate in vacuum
(to about 40--45\%\ of the free decay rate for the $s$--channel
and to about 30--35\%\ for the $p$--channel).

In their calculation, Dalitz and von Hippel~\cite{Dal64} found
that, compared to $\pi^-$ decay, Pauli suppression is about two
times stronger for $\pi^+$ decay of $^4_\Lambda$He. The reason is
that this $\pi^+$ decay, requiring both the $\Lambda$ and a proton,
happens inside the ``nuclear core,'' whereas the $\pi^-$
decay, involving only the $\Lambda$ itself, takes place outside
the core. We will use this value to estimate the relative
importance of Pauli suppression. We note, however, that Coulomb
repulsion (not considered by Dalitz and von Hippel) favors
$\pi^+$ decay, with three neutrons in the final state, over
$\pi^-$ decay, with three protons in the final state.

The transition $\Sigma^++N\to\pi^++n+N$
is assumed to be an $s$--wave three--body decay.
Apart from the fact that this is indicated by the experimental
$\pi^+$ energy distribution, as discussed above, there exists
additional evidence for this picture. In the hypertriton, the
virtual $\Sigma$ is found very close to one of the two nucleons,
see Fig.\ 5 of Ref.~\cite{Glo95}. The $\Sigma^+\,N$ pair
forms a tightly bound system, with binding energy of 80 MeV. 
The strong correlation of the $\Sigma^+$ with a nucleon indicates 
that the relevant decay is of the $\Sigma^++N\to\pi^++n+N$ three--body
type. We assume that, apart from the reduction in phase space,
the $\Sigma^+$ decay rate is unmodified in the medium; that
is, the $\Sigma^+$ in--medium three--body decay rate is taken
to be approximately equal to 
({\it i.e.}, to essentially replace)
the two--body free decay rate,
except for the phase space difference due to the $\Sigma^+$
being highly virtual.

The relevant decay ratio in vacuum which we need is
\begin{equation}
    \Gamma(\Sigma^+\to\pi^++n)\, / \,\Gamma(\Lambda\to\pi^-+p)
    = \frac{1}{2}\Gamma(\Sigma^+)\, / \,\frac{2}{3}\Gamma(\Lambda)
      \simeq 2.5 \ ,
\end{equation}
where the $\Delta I=1/2$ rule was used, which is well satisfied
experimentally.
Phase space gives an additional factor $70/185$, being the ratio
of the average $\pi^+$ momentum for in--medium $\pi^+$ decay of
$^4_\Lambda$He to its value for $\Sigma^+\to\pi^++n$ decay in
vacuum. Collecting factors, we estimate for the contribution of
our three--body virtual $\Sigma^+$ decay to the $\pi^+$ to $\pi^-$
branching ratio the value
\begin{equation}
  R(\pi^+/\pi^-) \simeq \frac{1}{2} \times
                 \frac{P(\Sigma^+)}{1-P(\Sigma^+)} \times 2.5 \times
                 \frac{70}{185} = 5(3)\% \ ,
\end{equation}
where the factor $1/2$ is due to the relative Pauli suppression 
discussed above, and we assume that the decay via the virtual 
$\Sigma^0$ component is counted primarily in the $\pi^-$ decay rate. 
The main uncertainty comes from that in the $\Sigma^+$ probability 
$P(\Sigma^+)$.

Our final value $R(\pi^+/\pi^-)=5(3)\%$ is to be compared to
the estimate 0.2\%\ (or 0.4\%\ with the higher average $\pi^+$
momentum) given by von Hippel~\cite{Hip64}, which is more than
an order of magnitude smaller.  We suggest that {\it a $\pi^+$ to
$\pi^-$ branching ratio $R(\pi^+/\pi^-)$ of the order of 5\% is a 
plausible result for any model calculation that includes $\Lambda-\Sigma$
conversion\/}.  Clearly, a more realistic model calculation which
includes the charge--exchange channel~\cite{Cie97} as well as the
$\Sigma^0+p\to\pi^++n+n$ decay mechanism is called for.

An additional reason why $\pi^+$ emission is less favorable from 
the light hypernuclei $^4_\Lambda$H and $^5_\Lambda$He
(compared to $^4_\Lambda$He) is that the resulting final states,
with four identical neutrons, are strongly suppressed by
the Pauli principle. Moreover, in the case of $^5_\Lambda$He,
breakup of the $^4$He core is required.
In contrast, $\pi^-$ decay for $^4_\Lambda$H and $^5_\Lambda$He
can give four nucleons in the favored $T=0$ $^4$He configuration.
However, the main reason that $\pi^+$ emission is not observed
is the absence of a significant $\Sigma^+$ admixture, which is
related to lack of an excess of positive charge in the system.

Finally, in order for our picture to hold, one should address
the absence of low--energy pions in the $\pi^-$ decay of 
$^4_\Lambda$He, shown in Fig.\ 1$a$. Two--body decay in process 
($a$) is by far the dominant decay mode; observation of three--body 
contributions would require a much larger data sample to see the few 
low--energy $\pi^-$s anticipated from the few \% decay branch of 
the $\Sigma^0$ or other three--body decay mechanisms .  The 
$^4_\Lambda$He has no virtual $\Sigma^-$ component to provide 
$\pi^-$s; in contrast, the $^4_\Lambda$H wave function is, 
schematically,
\begin{equation}
   |\,^4_\Lambda{\rm H}\rangle =
 \alpha\,|\Lambda\!\otimes\!\,^3{\rm H}\rangle + \beta\,(
   \sqrt{\frac{1}{3}}\,|\Sigma^0\!\otimes\!\,^3{\rm H}\rangle
  -\sqrt{\frac{2}{3}}\,|\Sigma^-\!\otimes\!\,^3{\rm He}\rangle ) \ ,
\end{equation}
with $P(\Sigma^-)\simeq 9\%$.  For $\pi^-$ decay of $^4_\Lambda$H 
low--energy pions are expected from the reaction 
$\Sigma^-+p\to\pi^-+n+p$ following the $\Lambda+n\to\Sigma^-+p$ 
virtual transition, but 5\% is still a small branch to observe.  
However, the $\pi^-$ decay of $^4_\Lambda$H is dominated by the 
analog transition to $\pi^- +\,^4$He, which absorbs some 70\% of 
the decay strength~\cite{Dav91}. That is, the analog transition 
acts as a two--body decay filter.  The remaining 30\% of the 
$\pi^-$ spectrum, which corresponds to the data in Fig.\ 1$c$ and 
1$d$, is sufficiently small that the $\sim$ 5\% of the $\pi^-$ 
decays coming from virtual $\Sigma^-$ and other three--body decay
mechanisms should be large enough to be seen. Indeed, a few
low--energy $\pi^-$ events are observed.

\section{SUMMARY AND CONCLUSIONS}
We have presented a plausible solution to the long--standing enigma
of $\pi^+$ emission in the weak decay of $^4_\Lambda$He. In the
spectrum of $\pi^+$s, one ``sees'' the weak three--body decay
of a highly--virtual $\Sigma^+$ arising from the in--medium
$\Lambda\leftrightarrow\Sigma$ transition. This resolution of
the $\pi^+$ decay puzzle is consistent with the significant
$\Sigma^+$ probability unique to the $^4_\Lambda$He hypernucleus,
and also explains in a natural way the experimentally observed
$\pi^+$ energy spectrum and $s$--wave angular distribution.
Moreover, the $\pi^+$ emission in $^4_\Lambda$He decay provides
direct confirmation of the important role played by virtual
transitions among members of the baryon octet, {\it in casu}
$\Lambda-\Sigma$ coupling.  This is additionally exemplified by
$\Lambda\Lambda-\Xi N$ coupling in $\Lambda\Lambda$
hypernuclei~\cite{Gib94,Car96}. As a final example, we mention
that it has been pointed out that signatures of $\Lambda-\Sigma$
mixing should also be visible in the magnetic moments of some
hypernuclei~\cite{Dov95}.

Based on our picture of a three--body decay of the virtual $\Sigma^+$
replacing two-body decay which one sees in the case of the $\Lambda$,
we conclude that the properties of the $\Sigma^+\to\pi^++n$ 
free decay are unrelated to the $^4_\Lambda$He $\pi^+$ decay 
observations.  Therefore, we disagree with previously published
conclusions that the fact that free $\Sigma^+$ decay is $p$--wave 
prevents the $\Lambda+p\to\Sigma^++n$ transition, followed by a 
$\Sigma^++N\to\pi^++n+N$ decay, from explaining the $\pi^+$ decay 
puzzle.

\acknowledgements
The authors wish to acknowledge the key role that Carl Dover played
in introducing them to the question.  Also, they thank A. Gal for
discussions relating to the role of the triton final state in the
$\pi^+$ decay.  The work of BFG was performed under the auspices of 
the U.\ S.\ Department of Energy. The work of RT was included in the 
research program of the Stichting voor Fundamenteel Onderzoek der 
Materie (FOM) with financial support from the Nederlandse Organisatie 
voor Wetenschappelijk Onderzoek (NWO).


\pagebreak

\begin{figure}
\caption{The $\pi^\pm$ kinetic energy distribution from the
decays: ($a$) $^4_\Lambda$He $\to\pi^-+p+\,^3$He, ($b$) $^4_\Lambda$He 
$\to\pi^++n+\,^3$H, ($c$) $^4_\Lambda$H $\to\pi^-+n+\,^3$He, and ($d$)
$^4_\Lambda$H $\to\pi^-+p+\,^3$H, as reproduced from
Ref.~\protect\cite{Sac66}.}
\end{figure}

\vspace{12pt}
\begin{figure}
\caption{The $\pi^+$ kinetic energy spectrum for all
uniquely identified decays of $^4_\Lambda$He observed in emulsion.
Reproduced from Ref.~\protect\cite{Boh69}.}
\end{figure}


\begin{thebibliography}{10}
\bibitem{May66} C.\ Mayeur, J.\ Sacton, P.\ Vilain, G.\ Wilquet, D.\
   O'Sullivan, D.\ Stanley, P.\ Allen, D.\ H.\ Davis, E.\ R.\ Fletcher,
   D.\ A.\ Garbutt, J.\ E.\ Allen, V.\ A.\ Bull, A.\ P.\ Conway, and
   P.\ V.\ March, Nuovo Cim. {\bf 44}, 698 (1966).
\bibitem{Key76} G.\ Keyes, J.\ Sacton, J.\ H.\ Wickens, and M.\ M.\
   Block, Nuovo Cim. {\bf 31A}, 401 (1976).
\bibitem{Afn90} I.\ R.\ Afnan and B.\ F.\ Gibson, Phys. Rev. C {\bf 40}, 
   R7 (1989); {\it ibid}. {\bf 41}, 2787 (1990).
\bibitem{Glo95} K.\ Miyagawa, H.\ Kamada, W.\ Gl\"ockle, and V.\ Stoks,
   Phys. Rev. C {\bf 51}, 2905 (1995).
\bibitem{Bod66} A.\ R.\ Bodmer, Phys.\ Rev.\ {\bf 141}, 1387 (1966).
\bibitem{Her67} R.\ C.\ Herndon, Y.\ C.\ Tang, Phys.\ Rev.\
   {\bf 153}, 1091 (1967); {\it ibid}. {\bf 159}, 853 (1967); {\bf 165},
   1093 (1969); R.\ H.\ Dalitz, R.\ C.\ Herndon, and Y.\ C.\ Tang, Nucl.\ 
   Phys.\ {\bf B47}, 109 (1972).
\bibitem{Gib72} B.\ F.\ Gibson, A.\ Goldberg, and M.\ S.\ Weiss,
   Phys.\ Rev.\ C {\bf 6}, 741 (1972);  B.\ F.\ Gibson, A.\ Goldberg, 
   and M.\ S.\ Weiss, in {\it Few Particle Problems in Nuclear 
   Interactions} (North Holland, Amsterdam, 1972), p.\ 188.
\bibitem{Gal75} A.\ Gal, Adv. Nucl. Phys. {\bf 8}, 1 (1975).
\bibitem{Jur73} M.\ Juric, G.\ Bohm, J.\ Klabuhn, U.\ Krecker,
   F.\ Wysotzki, G.\ Coremans--Bertand, J.\ Sacton, G.\ Wilquet,
   T.\ Cantrell, F. Esmael, A.\ Montwill, D.\ H.\ Davis,
   Nucl.\ Phys.\ {\bf B52}, 1 (1973).
\bibitem{Dav67} D.\ H.\ Davis and J.\ Sacton, in {\it High Energy Physics}, 
   Vol. II (Academic Press, New York, 1967) p. 365.
\bibitem{Dav91} D.\ H.\ Davis, in the Proceedings of the {\it LAMPF
   Workshop on ($\pi,K$) Physics}, A.I.P.\ Conf.\ Proc.\ {\bf 224},
   edited by B.\ F.\ Gibson, W.\ R.\ Gibbs, and M.\ B.\ Johnson
   (American Institute of Physics, New York, 1991), p. 38.
\bibitem{Dab73} J.\ Dabrowski and E.\ Fedorynska, Nucl. Phys.
   {\bf A210}, 509 (1973).
\bibitem{Gib75} B.\ F.\ Gibson and D.\ R.\ Lehman, Phys. Rev. C
   {\bf 11}, 29 (1975).
\bibitem{Dal62} R.\ H.\ Dalitz and G.\ Rajasekharan, Phys. Lett.
   {\bf 1}, 58 (1962).
\bibitem{Dal64} R.\ H.\ Dalitz and F.\ von Hippel, Nuovo Cim.
   {\bf 34}, 799 (1964).
\bibitem{Hip64} F.\ von Hippel, Phys.\ Rev. {\bf 136}, B455 (1964).
\bibitem{Dal66} R.\ H.\ Dalitz, in
   the Proceedings of the International School 
   of Physics ``Enrico Fermi'' XXXVIII, {\it Interaction of High--Energy
   Particles with Nuclei}, edited by T.\ E.\ O.\ Ericson (Academic Press,
   New York--London, 1967), p. 89.
\bibitem{Cie97} A. Ciepl\'y and A. Gal, Phys.\ Rev.\ C {\bf 55}, 2715
   (1997).
\bibitem{Boh69} G.\ Bohm, J.\ Klabuhn, H. Krecker, F. Wysotzki, G.\
   Coremans, W. Gajewski, C.\ Mayeur, J.\ Sacton, P.\ Vilain, G.\ Wilquet,
   D. O'Sullivan, D.\ Stanley, D.\ H.\ Davis, E. R. Fletcher, S.\ P.\
   Lovell, N.\ C.\ Roy, J.\ H.\ Wickens, A.\ Filipkowski, K.\
   Garbowska--Pniewska, T. Pniewski, E.\ Skrzypczak, T.\ Sobczak, J.\ E.\
   Allen, V.\ A.\ Bull, A.\ P.\ Conway, A.\ Fishwick, and P.\ V.\ March,
   Nucl.\ Phys. {\bf B9}, 1 (1969).
\bibitem{Sac66} J. Sacton, in Ref.~\cite{Dal66}, p. 77.
\bibitem{Gaj66} W. Gajewski, J.\ Sacton, P.\ Vilain, G. Wilquet,
   D.\ Stanley, D.\ H.\ Davis, E.\ R.\ Fletcher, J.\ E.\ Allen, V.\ A.\
   Bull, A.\ P.\ Conway, and P.\ V.\ March, Phys.\ Lett. {\bf 21}, 
   673 (1966).
\bibitem{Kre74} U.\ Krecker, D.\ Kielczwska, and T.\ Tymieniecka,
   Nucl.\ Phys. {\bf A236}, 491 (1974).
\bibitem{Ada73} O.\ Adamovic, U.\ Krecker, G.\ Coremans--Bertrand, J.\
   Sacton, T.\ Cantwell, C. Ni Ghogain, D.\ H.\ Davis, D.\
   Kielczewska, T.\ Tymieniecka, and Z.\ Zakrzekski, Lett.\ Nuovo
   Cim. {\bf 6}, 9 (1973).
\bibitem{Dal59} R.\ H.\ Dalitz and L.\ Liu,
   Phys. Rev. {\bf 116}, 1312 (1959).
\bibitem{Gib94} B.\ F.\ Gibson, I.\ R.\ Afnan, J.\ A.\ Carlson,
   and D.\ R.\ Lehman, Prog. Theor. Phys. Suppl. {\bf 117}, 339 (1994).
\bibitem{Car96} S. B. Carr, I. R. Afnan, and B. F. Gibson,
   Nucl. Phys. {\bf A} (accepted for publication).
\bibitem{Dov95} C.\ B.\ Dover, H.\ Feshbach, and A.\ Gal,
   Phys. Rev. C {\bf 51}, 541 (1995).
\end{thebibliography}
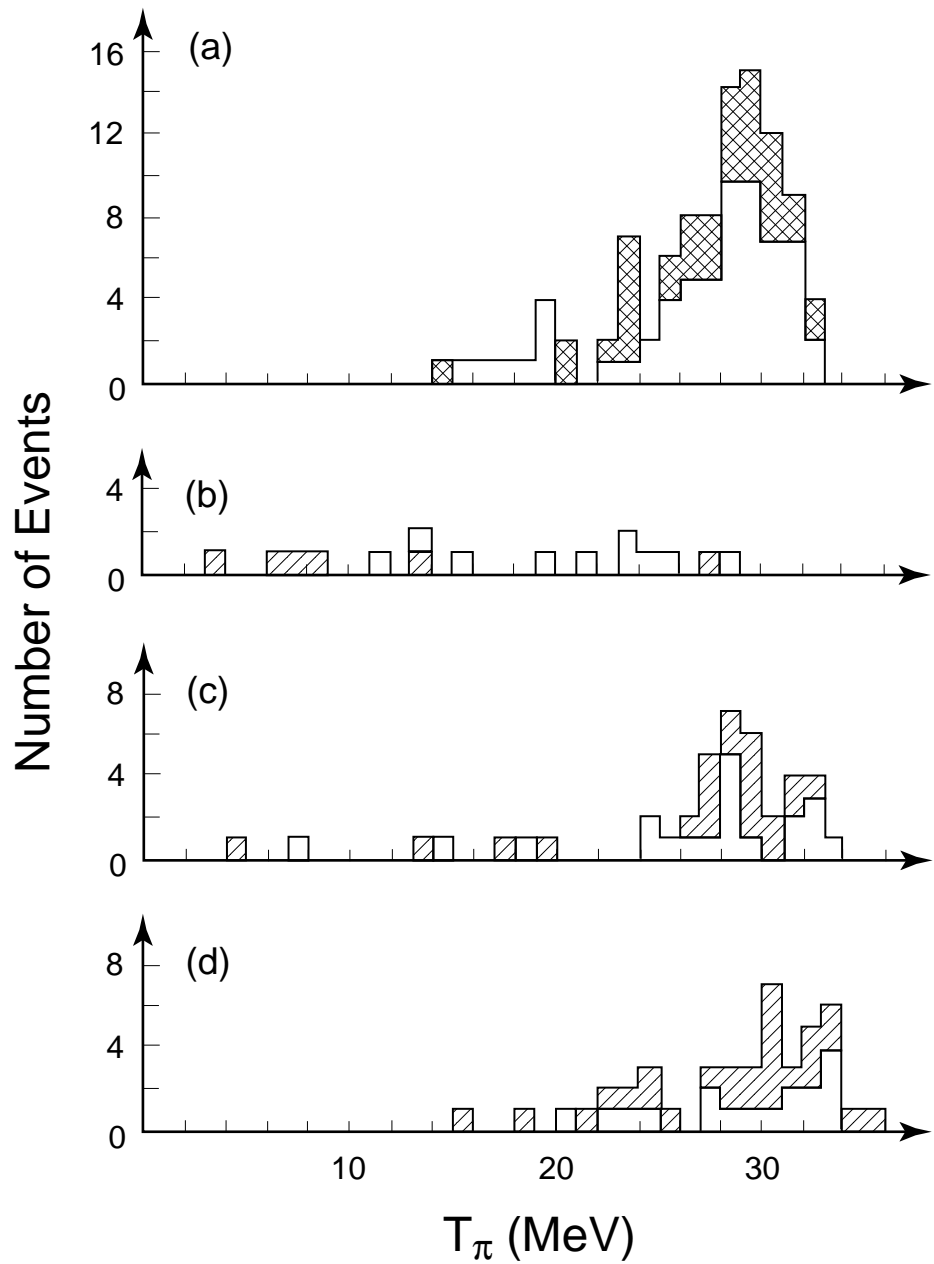
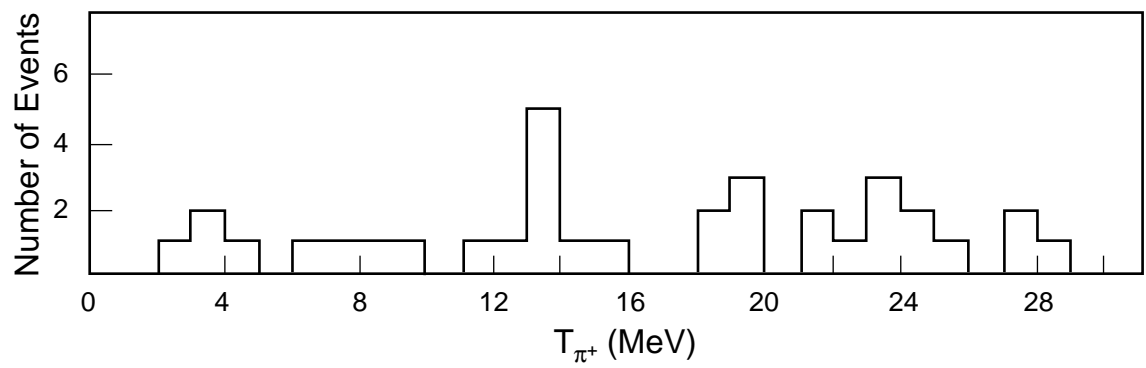
\end{document}